\documentstyle[12pt,epsfig]{article}
\title{Cosmic ray spectrum and anisotropies\\from the knee to the second knee}
\author{Juli\'an Candia$^a$, Silvia Mollerach$^b$ and Esteban Roulet$^b$\\
$^a${\it Departamento de F\'{\i}sica, Universidad Nacional de La Plata, 
CC67,}\\{\it La Plata 1900, Argentina}\\
$^b$ {\it CONICET, Centro At\'omico Bariloche, Av. Bustillo 9500,}\\
{\it Bariloche 8400, Argentina}}

\begin{document}
\maketitle

\begin{abstract}
We consider the scenario in which the knee in the cosmic ray spectrum 
is due to a change in the escape mechanism of cosmic rays from the Galaxy 
from one dominated by transverse diffusion to one dominated by drifts. 
We show that this scenario explains not only the changes in spectral 
slope at the knee and at the second knee, but can also account for the 
main characteristics of the observed energy dependent anisotropy amplitude 
and phase of first harmonic in the energy range between $10^{15}$ and
$10^{18}$~eV. This provides a useful handle to distinguish this diffusion/drift
model from other scenarios proposed to explain the knee in the spectrum.   
\end{abstract}
\section{Introduction}
A puzzling feature of the cosmic ray (CR) spectrum is the so-called knee,
i.e. a steepening taking place at an energy 
$E_{knee}\simeq 3\times10^{15}$~eV.   
Several models have been proposed so far in order to explain
this feature, although none of them has managed to become broadly accepted.
Some proposals focus on a possible crossover between different
acceleration mechanisms below and above the knee \cite{la83,bi93,dr94}, 
or exploit the possibility of a change in the
particle acceleration efficiency \cite{fi86,jo86,ko00}.  
Other scenarios include the effects of nuclear photodisintegration processes at the
sources \cite{ka93,ca02a}, the recent explosion of a single 
source \cite{er97}, and increased leakage from the Galaxy due
to a change in the confinement efficiency of CRs by galactic 
magnetic fields \cite{sy71,wd84,pt93,ca02b}.
However, considering the information on the spectrum alone 
it is difficult to discriminate among the predictions worked out from 
the proposed models, and so other kind of observations, 
namely the CR mass composition and the anisotropy
measurements, are decisive in testing the different proposals. 

For instance, the scenario that considers nuclear 
photodisintegration at the sources predicts a CR composition 
that becomes lighter above the knee, 
while other (rigidity dependent) explanations 
predict the opposite trend. Unfortunately, the observational status is far
from being clear in this respect \cite{sw02}, with some observations 
\cite{sw00} suggesting that the CR composition could turn lighter while others
finding that the heavier components become dominant \cite{ka99}. 

On the other hand, anisotropy observations around and above the knee appear
in this context as a valuable means of imposing further  
restrictions on the possible scenarios. In particular, scenarios in which the knee
is due to a change in the acceleration efficiency 
at the sources \cite{fi86,jo86,ko00}, or to the
effects of nuclear photodisintegration in 
them \cite{ka93,ca02a}, are not expected to 
lead to any particular change in the behavior of the observed anisotropies in
correlation with the observed change in the spectral index at the knee energy,
while scenarios in which the knee is linked to a change in the efficiency of
CR confinement in the Galaxy \cite{sy71,wd84,pt93,ca02b} 
can predict such correlation. Moreover, to account for
the observed energy dependence of the anisotropy amplitude and phase 
of first harmonic in the whole range of energies between the knee and the ankle is
a highly non-trivial requirement for any of these scenarios. 

The aim of this work is to perform a more detailed comparison between the
anisotropy observations and the predictions within the scenario in which the knee is due
to a crossover from a regime in which the CR transport is
dominated by transverse diffusion to one dominated by drift effects \cite{pt93},  
recently reconsidered in detail in refs. \cite{ca02b,ca02c}, and hereafter referred
as the diffusion/drift model. This analysis will be done for the whole
range of energies in which drift effects are relevant ($E>E_{knee}$) and where CRs 
are dominated by the galactic component ($E<E_{ankle}\simeq5\times10^{18}$~eV). 
The CR transport in the Galaxy is determined by solving numerically a diffusion
equation that takes account of the regular galactic magnetic field as well as a
random turbulent field with a Kolmogorov spectrum of inhomogeneities, 
while for the extragalactic component an isotropic flux
is adopted. Then, under different assumptions concerning the galactic field model considered, 
the particular set of parameters adopted within the given model (i.e. the field amplitudes
and scale heights assumed for both the regular and random field components), the source
distribution and the spectrum of the extragalactic (isotropic) flux, the results are contrasted to
the observed CR energy spectrum, the anisotropy amplitude and phase of first harmonic measured
by different experiments. It is found that the results show the right tendency 
to naturally account for the observed features,
and hence these observations provide further support to this model. 
\section{The diffusion/drift model}
To understand the main physical ingredients of the diffusion/drift model considered in
refs. \cite{pt93,ca02b,ca02c}, 
let us assume the CR transport in the Galaxy to be described by 
a steady-state diffusion equation 
 ${\bf \nabla}\cdot{\bf J} = Q$, where $Q$ is the source and
the CR current is related to the CR density $N$ through
\begin{equation}
{\bf J}=-D_\perp {\bf \nabla}_\perp N- D_{\parallel} 
{\bf \nabla}_{\parallel} N
+D_A {\bf b}\times{\bf \nabla}N\ \ ,
\label{current}
\end{equation}
with ${\bf b}$ being the unit vector in the direction of the regular
magnetic field ${\bf B}_0$, i.e. ${\bf b}\equiv 
{\bf B}_0/|{\bf B}_0|$, and 
${\bf \nabla}_\parallel={\bf b}({\bf b}\cdot
{\bf \nabla})$, while ${\bf \nabla}_\perp={\bf \nabla}-
{\bf \nabla}_\parallel$. The components of the diffusion tensor are 
$D_{\parallel}$ (along the direction of ${\bf B}_0$), 
$D_\perp$ in the
perpendicular direction, while $D_A$ is associated to the
antisymmetric part and determines the drift effects. Assuming for
simplicity that the regular magnetic field is directed in the
azimuthal direction and that in a first approach both the Galaxy 
and the CR sources can be
considered to have cylindrical symmetry, one finds that $D_\parallel$ 
plays no role in the diffusion equation,
which involves then only $D_\perp$ and $D_A$. These diffusion coefficients
are obtained from the approximate expressions given in ref. \cite{ca02b}.
Indeed, defining the turbulence level $\eta\equiv B_{rand}^2/(B_{rand}^2+B_0^2)$, 
where $B_{rand}$ is the rms amplitude of the 
random component of the magnetic field,
the relevant diffusion coefficients are well described in the range of
interest of this work by
\begin{equation}
D_{\perp} \simeq 8.7\times 10^{27} e^{3.24\eta} 
\left(1-\eta\right)^{1/6}\left({{r_L}\over{\rm{pc}}}\right)^{1/3} 
{{\rm{cm^2}}\over{\rm{s}}}  
\label{dperp}
\end{equation}
and
\begin{equation}
D_A\simeq 3\times 10^{28}\left(1+0.049\left({{\eta}\over{1-\eta}}\right)^{3/2}
\right)^{-1}\left({{r_L}\over{\rm{pc}}}\right){{\rm{cm^2}}\over{\rm{s}}} \ \ ,  
\label{da}
\end{equation}
where $r_L$ is the CR Larmor radius. The spatial distribution of sources will be
specified below in Section 3.  

The change in the spectrum from the diffusion 
to the drift dominated regimes can be simply understood from the fact that,
as turns out from Eqs.(\ref{dperp}) and (\ref{da}), the energy dependence 
of the diffusion coefficients is given by
$D_\perp ({\bf x})\simeq D_\perp^0 ({\bf x}) (E/E_0)^{1/3}$ and
$D_A({\bf x}) \simeq D_A^0 ({\bf x}) E/E_0$ \cite{pt93}\footnote{In
general, for a spectrum of random magnetic field inhomogeneities 
satisfying ${\rm d}B^2/{\rm d}k\propto k^{-a}$ (where $a=5/3$ corresponds
to the Kolmogorov case, or $a=3/2$ for a Kraichnan \cite{kr65} hydromagnetic 
spectrum), one has $D_{\perp}\propto E^{2-a}$.}. Hence, 
one finds that at low energies ($E<Z~E_{knee}$, where $Z$ is the CR charge) 
the transverse 
diffusion is the dominant process affecting the transport of CRs and
leading to ${\rm d}N/{\rm d}E\propto(D_\perp^0/D_\perp){\rm d}Q/{\rm d}E\propto
E^{-\beta-1/3}$, where $\beta$ is the spectral index of the source,
while at high energies ($E\gg Z~E_{knee}$) one has instead 
${\rm d}N/{\rm d}E\propto(D_A^0/D_A){\rm d}Q/{\rm d}E\propto E^{-\beta-1}$.
Thus, we see that for a constant source spectral index $\beta\simeq 2.4$,
one has the correct spectral slope below the knee 
(${\rm d}N/{\rm d}E\propto E^{-\alpha}$, with $\alpha\simeq 2.7$). 
Each CR component of charge $Z$
starts to be affected by drifts at an energy $E\simeq Z~E_{knee}$, and its
spectrum progressively steepens, with the spectral index finally
changing by $\Delta\alpha\simeq 2/3$ in a decade of energy. The
envelope of the total spectrum obtained by adding together the
different nuclear components nicely fits the change from a spectrum
$\propto E^{-2.7}$ below the knee, to one $\propto E^{-3}$ above
it. Moreover, since all the lighter components are
strongly suppressed above $10^{17}$~eV, 
the dominant iron component will progressively
steepen its spectrum  until the overall spectrum becomes $\propto
E^{-2.7-2/3}$ above a few $\times 10^{17}$~eV, hence also
reproducing the behavior observed at the so-called second knee, namely
a second steepening taking place at $E_{sk}\simeq 4\times10^{17}$~eV,
in which the spectral slope is close to $\alpha\simeq 3.3$.   
Furthermore, we have shown \cite{ca02b} that this scenario predicts a CR mass 
composition which is compatible with those experiments that report a  
heavier composition above the knee.             
Concerning the anisotropy observations, we have also shown \cite{ca02c} that 
a few~$\%$ anisotropy with an excess from a direction near the 
galactic center and a deficit near the galactic anticenter 
direction around $\sim 10^{18}$~eV \cite{te01} could be accounted for by 
diffusion and drift currents affecting the galactic CR component.
In this way, this simple scenario that just takes into account well established 
properties of the propagation of charged particles in regular
and turbulent magnetic fields can provide an explanation of quite diverse 
galactic CR observations without requiring any additional assumptions.
\section{The cosmic ray spectrum and anisotropies}
The contribution to the anisotropy from galactic CRs of a given 
charge $Z_i$ will be \cite{be90}
\begin{equation}
\delta_i=\frac{3~{\bf J_i}}{c~N_i}\ \ ,
\label{anis}
\end{equation} 
where ${\bf J_i}$ is the CR current corresponding to the given component, and
is hence given by an expression 
analogous to  Eq.(\ref{current}) but involving the densities $N_i$ 
corresponding to the CRs with charge $Z_i$.
The total anisotropy will then be given by 
$\delta=\sum_if_i\delta_i$, where $f_i\equiv N_i/N$ 
are the fractional abundances of all CR species
(with $N$ the total, galactic plus extragalactic, CR density). 
If we assume that the extragalactic flux is isotropic, it can be shown that
this flux, which presumably gives the 
dominant contribution above the ankle, will not be enhanced by the diffusion process
and hence will remain isotropic \cite{ca02c,cl96}. 
This implies that the presence of the assumed 
isotropic extragalactic component just reduces the fractions $f_i$
of the galactic components, having the effect of suppressing the growth of 
the total anisotropy amplitude as the energy of the ankle is approached.   

Notice that under the assumptions of cylindrical symmetry and that the regular
magnetic field is in the azimuthal direction, one has that 
${\bf \nabla}_\perp N={\bf \nabla} N$ and also that the CR current will be
perpendicular to the regular magnetic field (i.e. lying on the $r-z$
plane). Moreover, the contribution to the CR current arising from the transverse 
diffusion will be in the direction of ${\bf\nabla} N$,
while the drift part will be orthogonal to ${\bf\nabla} N$. 

As we mentioned before, at low energies the diffusion is completely 
determined by $D_{\perp}$, and hence 
the contribution to the anisotropy from CRs of charge $Z_i$  
is $\delta_i \propto D_{\perp}\nabla N_i/N_i$, increasing with energy 
roughly as $\delta_i\propto E^{1/3}$. However, at higher energies
there is a crossover to the drift dominated regime, leading to the behavior 
$\delta_i\propto E$. At even higher energies, above $10^{18}$~eV, 
the CR density gradually  starts 
to be dominated by the extragalactic component, 
and hence the anisotropy will be determined by 
the intrinsic anisotropy of the extragalactic 
component (which for simplicity was assumed to be negligible in this work).
  
The field models considered in this work are similar to those studied in ref. 
\cite{ca02b} (except for a minor modification of the vertical profile function).
We will consider for simplicity that the regular magnetic field 
is directed in the azimuthal direction and that the system possesses azimuthal symmetry.
The propagation region of the galactic CRs will be taken as 
a cylinder of radius $R=20$~kpc and height $2H$, with a halo size $H=10$~kpc. 
Adopting cylindrical coordinates $(r,\phi,z)$ throughout, 
the regular galactic magnetic field can be expressed as
${\bf{B_0}}=(B_0^{disk}+B_0^{halo})\hat{\phi}$, where $B_0^{disk}$ and $B_0^{halo}$ are 
$\phi$-independent functions that correspond to the disk and halo regular field components,
respectively. The disk component is assumed to be given by
\begin{equation}
B_0^{disk}(r,z)=B_d\sqrt{{{1+(r_{obs}/r_c)^2}}\over{1+(r/r_c)^2}} 
\sin\left({{\pi r}\over{4~{\rm kpc}}}\right){\rm
th}^2\left(\frac{r}{1\ {\rm kpc}}\right){{1}\over{\cosh(z/z_d)}} \ \ ,
\label{b0d}
\end{equation}
where $r_c=4$~kpc is a core radius which smoothes out  near the
galactic center the overall $1/r$ behavior usually considered, 
$z_d$ is the disk scale height, 
and the value of $B_d$ is chosen such that 
$B_0^{disk}(r_{obs},z_{obs})=-1$~$\mu$G at our observation point 
($r_{obs}=8.5$~kpc, $z_{obs}=0$), the minus sign arising from the fact 
that the local galactic magnetic
field is nearly directed in the $-\hat{\phi}$ direction. 
The th$^2(r)$ factor has been introduced
just to insure that there are no singularities in the $r=0$ axis,
since otherwise a singular drift current along that axis would be 
artificially produced.  
Notice also that this regular field component has reversals in 
its direction for $r=4,$ 8, 12 and 16~kpc.

The halo component can be described either by a field structure with radial 
reversals and also symmetric with respect to the galactic
plane (denoted by $R$-$S$), or by a model without radial reversals (denoted $NR$) 
which may correspond to an independently generated halo field.
In the latter case four different configurations arise from choosing the relative 
orientation of the halo field relative to the disk one, 
and from the symmetry with respect to the galactic plane ($S$ models having 
$B(r,-z)=B(r,z)$, while $A$ models having $B(r,-z)=-B(r,z)$). In this work, 
we will consider only the $NR$-$A+$ and $NR$-$S+$ cases, 
the $+$ sign denoting that 
sign$(B_0^{halo}\cdot B_0^{disk})>0$ at our galactocentric radius for $z>0$,
since a detailed discussion comparing the different field configurations has already 
been worked out previously \cite{ca02b}. 

The halo component is then taken as 
\begin{equation}
B_0^{halo}(r,z)=B_h\sqrt{{1+(r_{obs}/r_c)^2}\over{1+(r/r_c)^2}} 
{\rm th}^2\left(\frac{r}{1\ {\rm kpc}}\right)\ {{1}\over{\cosh(z/z_h)}}
\ \ {\mathcal{R}} \ \ ,  
\label{b0h}
\end{equation}
where the normalization factor is assumed to be $B_h=-1\ \mu$G, 
and a halo scale height $z_h$ 
is introduced in the vertical profile function. 
The function $\mathcal{R}$
is just unity for the symmetric halo case, and it is taken as 
${\mathcal{R}}(z)={\rm tanh}(z/z_d)$ 
in the case of a halo model 
antisymmetric with respect to the galactic plane. 
The Faraday rotation measures of pulsars and extragalactic radio sources
might indicate that the global field structure is best described by
a bi-symmetric spiral field in the disk with reversed direction from
arm to arm, and an azimuthal field in the halo with reversed directions 
below and above the galactic plane, which could correspond to an A0
dynamo field configuration \cite{ha02}. In that case, the $NR$-$A+$ field 
model may reflect appropriately the main features of the large-scale 
galactic field structure. However, according to present observational data, 
the origin of the galactic field in the halo could also be related to the 
origin of the disk field, for instance if a galactic wind is present, somehow 
extending the field disk properties into the halo. Hence, the $NR$-$S+$ field model,
as well as other variants discussed in ref. \cite{ca02b}, are also plausible.  

For the random component, its intensity is taken as
\begin{equation}
B_{rand}(r,z)=B_r\sqrt{{1+(r_{obs}/r_c)^2}\over{1+(r/r_c)^2}} 
{{1}\over{\cosh(z/z_r)}} \ \ ,
\label{brand}
\end{equation}
where $z_r$ is the corresponding scale height and $B_r$ the local rms amplitude. 

The source term in the diffusion equation is assumed to take the form 
$Q(r,z)={\mathcal Q}(r)\theta(h_s-|z|)$, with $h_s=200$ pc, and where
the radial profile of the source distribution is considered to be either  
localized at a given radius $r_s$ (i.e. ${\mathcal Q}(r)=\delta(r-r_s)$), or as
being constant within an inner and an outer radius. 

Since we deal here with a rigidity-dependent scenario, the results obtained depend on $E/Z$
and a convolution of the contribution of each component of charge $Z$ has to be made  
summing over all components in order to get the all-particle prediction.
The source spectrum of a given CR component of charge $Z$ is assumed 
to have an energy dependence d$Q_Z/{\rm d}E\propto E^{-\beta_Z}$,
where the normalization factors and spectral indices were 
obtained  from the experimental data
at low energies below the knee \cite{wi98,ho02}, 
which provide the local observed spectral indices $\alpha_Z=\beta_Z+1/3$,  
and extrapolated to higher energies 
without considering upper cutoffs. Also the relative abundances of the 
different elements were obtained from the same fits to the low energy 
observed CR densities \cite{wi98,ho02}.
In addition, an isotropic extragalactic component given by
\begin{equation}
\left({{\rm{d}N}\over{\rm{d}E}}\right)_{XG}=
1.7\times 10^{-33}\ \left({{E}\over{10^{19}{\rm eV}}}\right)^{-2.4}
{\rm m^{-2}s^{-1}sr^{-1}eV^{-1}}  
\label{xgflux}
\end{equation}
is added to the galactic flux as well. The energy dependence assumed for the extragalactic
CR contribution is similar to that inferred for the production at the sources 
(i.e. $\beta\simeq 2.4$), since diffusion
effects cannot enhance this flux. The amplitude of the extragalactic component is normalized 
to fit the observed fluxes for $E\simeq 10^{19}$~eV. 
 
In figure 1, the CR spectra and anisotropy amplitudes 
obtained for both the $NR$-$A+$ and $NR$-$S+$ field models are compared to the 
observations from several experiments 
\cite{na00,hi84,fi86}\footnote{The anisotropy experimental data
points are obtained from the amplitude of the first harmonic
in right ascension $A$, which is related to the anisotropy
$\delta$ through $A=\delta\cos d\cos\lambda$, where $d$ is
declination at which the observations are performed and
$\lambda$ is the latitude of the direction where the flux is maximum. 
The plotted points correspond to $A/\cos d$ for the different experiments,
and thus are actually a lower limit for $\delta$, as $\cos\lambda$
cannot be recovered from right ascension harmonic analysis.}.  
The set of field parameters ($B_r,z_d,z_h,z_r$) used for each model 
is indicated in the figure in $\mu$G and kpc respectively. 
The effects of varying either the field
model parameters or the source distribution have already been discussed in 
detail in ref. \cite{ca02b}. Very briefly, a turbulence increase, 
which can be performed either by increasing the random
field rms amplitude $B_r$ or by increasing the random field scale height $z_r$,
or also by decreasing the vertical scale heights $z_d$ and $z_h$ of the regular components, 
tends to suppress the drift effects, thus yielding larger fluxes and shifting
the bent in the spectrum to larger energies. 
\begin{figure}
\centerline{{\epsfxsize=5.5truein \epsffile{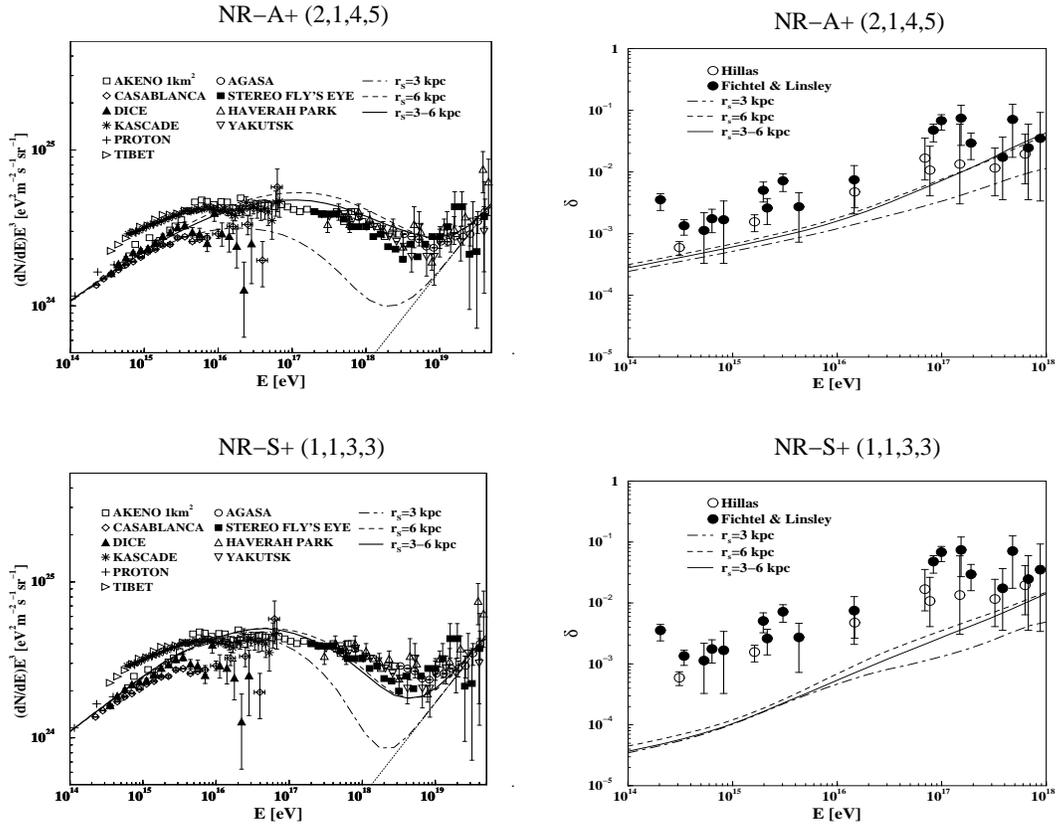}}}
\caption{CR spectra and anisotropy amplitudes 
computed for different galactic magnetic field
models and defining parameters, which are indicated as 
($B_r,z_d,z_h,z_r$) in $\mu$G and kpc respectively.    
The dotted straight lines correspond to the extragalactic 
flux given by Eq.(\ref{xgflux}).
Also shown are the relevant experimental data points.}
\label{fig1}
\end{figure}
On the other hand, the drift effects are stronger the farther the source is located, 
since the drift currents can remove the propagating CRs from the galactic plane
all the way from the source to the observer. Moreover, extended source distributions
tend to produce flatter spectra due to the smaller associated density gradients.
The plots of figure 1 also show that, when considering an extended ring of sources
with constant strength, the dominant observed flux is produced near the border 
which is closest to the observation point.  

~From figure 1 we observe that, assuming quite plausible field model parameters, 
the diffusion/drift scenario is capable of reproducing 
remarkably well not only the observed CR spectrum, but also the anisotropy 
measurements. Notice from Eqs.(\ref{current}) and (\ref{anis}) that $|\nabla N|$ plays a crucial
role in the enhancement of the anisotropy amplitude, and thus it turns out that 
close by sources produce larger anisotropies as compared to farther ones.
The effect of considering extended source distributions may be 
thought as given by the sum of the results produced by 
localized sources, weighted by the relative contribution of each 
one to the total CR density observed. Hence, it can happen that
just a small region within the extended distribution gives the
overwhelming contributions to the total CR flux and anisotropy
observed. However, it can also happen that different regions in 
the source distribution contribute with CR drift currents directed
along very different directions and comparable in magnitude, giving 
some regions of suppressed anisotropy, while producing anisotropy
enhancements in others. Anyhow, the results shown for 
localized sources should be indicative of the actual trend that
would be obtained for different source distributions, as follows from figure 1.    

Although some results for the $NR$-$S+$ case exhibit somewhat low anisotropies, 
it should be noticed that,
even in the case of localized sources, we are actually dealing with rings around the
galactic center, since non-central point-like sources would break the azimuthal 
symmetry assumed. Removing this simplifying assumption and solving the CR diffusion
equation for a point-like nearby source, a further increase in the anisotropy
amplitude might be achieved. Indeed, it has been recognized that the effect
of discrete nearby sources, as for instance pulsars and supernova remnants 
such as Vela, Loop III and Geminga \cite{do84,be90},
could actually give the dominant contribution to the observed anisotropies 
at low energies below the knee. 
The precise estimate of this effect requires however the knowledge of the spatial distribution, 
power, age, and evolution of the sources.
Hence, an accurate fit of our results to the experimental data below the knee is not 
really necessary, but it is reassuring to observe that the predictions  
within this scenario tend
to exhibit a quite acceptable agreement with observations above the knee.

In performing the sum over all components in order to obtain the all-particle
CR flux, we have taken into account the contribution of all nuclear species from
hydrogen to nickel. 
We have actually also explored
the possible contribution of heavier elements up to uranium, following ref. \cite{ho02}.
While the spectral indices of the elements with $Z\leq 28$ are measured at low energies
(below $\sim 10^{14}$~eV), and then the inferred source index $\beta_Z$ is simply
extrapolated to higher energies assuming that it remains constant,   
there is no information available on the spectral indices corresponding to
ultra-heavy elements (i.e., with $Z>28$). 
It was proposed in ref. \cite{ho02} to obtain them 
just extrapolating the possible $Z$-dependence of the observed 
spectral indices of the lighter elements, 
using a parametrization given by $-\alpha_Z=A+BZ^C$ to fit the data. 
While this three-parameter
expression favored a non-linear extrapolation (with $C=1.51\pm0.13$), 
also the case of a linear 
extrapolation (i.e, $C\equiv 1$) was considered and found 
to fit well the measured spectral indices \cite{ho02}. 
Assuming the linear extrapolation of spectral indices, we found that the
contribution of ultra-heavy elements is completely negligible 
within the diffusion/drift
scenario. These ultra-heavy elements would only have a noticeable effect
with the extreme non-linear extrapolations (for which for instance the 
spectral index for uranium results $\alpha_U\simeq 1.9$), but in our
model this could not lead to a sufficient suppression of the spectrum 
above $10^{17}$~eV,
unless one adopts an upper rigidity cutoff in the source near $E\simeq Z\times 10^{17}$~eV.
In ref. \cite{ho02} instead, an ad hoc steepening at the knee of $\Delta\alpha\simeq 2$
was adopted, and the resulting suppression was actually too large above the second knee.
One could also mention that the most natural expectation would be to have an universal
rigidity source spectrum, and hence the differences in the spectral indices observed at low
energies most likely would reflect the effects of spallation processes affecting heavy nuclei.
In this case no particular enhancements of ultra-heavy nuclei above the knee should be expected. 

\begin{figure}
\centerline{{\epsfxsize=3truein \epsffile{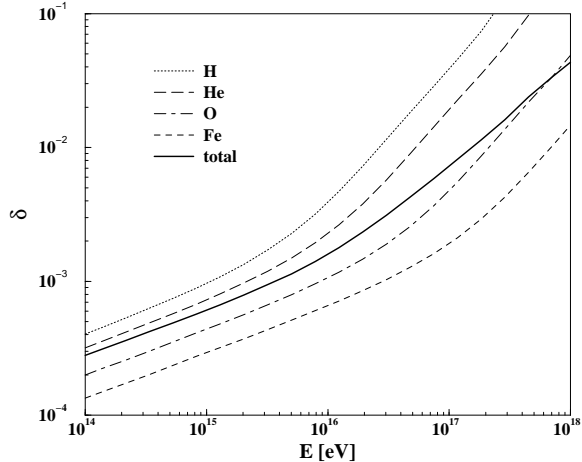}}}
\caption{Anisotropy amplitudes corresponding to the main galactic CR components, 
namely protons and nuclei of helium, oxygen and iron. 
For comparison, also the total (galactic plus extragalactic) anisotropy
amplitude is shown.  
The data correspond to the $r_s=3-6$~kpc extended source ring, the $NR$-$A+$ case and
the same field parameters as in figure 1.}
\label{fig2}
\end{figure} 

The anisotropy amplitudes of different galactic CR components are plotted in figure 2 as 
functions of the energy, for the $r_s=3-6$~kpc extended source ring, the $NR$-$A+$ case and
the same field parameters as before. As expected within the 
diffusion/drift scenario, the onset of the drift-dominated regime 
is a rigidity-dependent phenomenon that increases the anisotropy of all 
galactic CR components. 
It should be pointed out that, although light elements (particularly protons and He nuclei) 
are strongly suppressed at high energies due to leakage effects, they still contribute 
significantly to the total anisotropy. For instance, 
He nuclei are found to contribute around $30-40\%$ of the total galactic CR anisotropy 
in the whole energy range between $10^{14}$ and $10^{18}$~eV for the case shown in figure 2,
even though its abundance falls below $10\%$ at $10^{18}$~eV.   
It would certainly be of great interest to resolve
observationally the anisotropies of separate CR components around and above the knee, 
since that could help discriminate among some of the models proposed so far.   
Indeed, KASCADE may actually be capable of performing this kind of observations
in the near future\footnote{We thank Ralph Engel for bringing this issue to
our attention.}.

In figure 3 we display the anisotropy phase of first harmonic
in right ascension vs. energy, 
again compared to the experimental observations 
compiled in refs. \cite{hi84,fi86}, for the same cases considered in figure 1. 
At low energies, the transverse diffusion dominates and the direction of maximum 
intensity corresponds to the galactic center, as expected \cite{ca02c}.
At higher energies the phase of the anisotropy depends actually on the detailed
geometry of the regular field adopted, but in the models considered the differences
are not large. However, as shown in ref. \cite{ca02c}, for other magnetic field
models the anisotropy maximum may point towards the galactic center near $10^{18}~$eV. 

\begin{figure}
\centerline{{\epsfxsize=3truein \epsffile{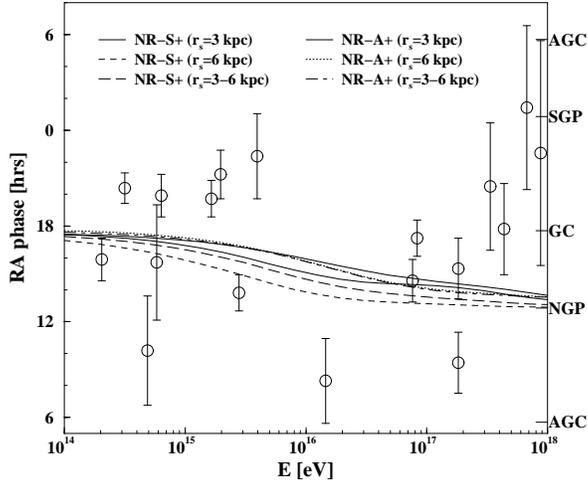}}}
\caption{Anisotropy phase of first harmonic
in right ascension vs. energy obtained for the same cases as in figure 1.
For comparison, experimental data
measured by different experiments is also shown.
The labels on the right indicate the directions of maximum CR intensity
(AGC: Anti Galactic Center, NGP: North Galactic Pole, etc.), taking into
account that $\delta$ is contained in the $r$-$z$ plane.} 
\label{fig3}
\end{figure}

As a summary, in this work we considered the diffusion/drift model to study the 
propagation of galactic CRs. In agreement with previous results
\cite{ca02b,ca02c}, we find that this scenario explains
remarkably well the observed spectrum from below the knee up to
the ankle, and in particular the changes in
spectral slope at the knee and at the second knee. Furthermore,   
from the very nature of this scenario, 
the anisotropy amplitude naturally experiences a crossover 
from the transverse diffusion dominated regime 
to the drift dominated one, which shows up around the knee region.    
Indeed, this is a particular feature that very clearly 
differentiates this diffusion/drift scenario from other proposals in which such a change in
the energy dependence of the anisotropy amplitude is not to be expected. 
Recalling some well known statements claiming a kind of correlation 
between the spectrum steepening at the knee and the occurrence of an anisotropy enhancement 
\cite{hi84}, one could hence expect anisotropy considerations to favor 
this kind of scenarios against other proposals.  
Hence, more detailed and precise measurements concerning anisotropy observations will 
certainly be decisive in order to test this diffusion/drift scenario,
as well as the other models proposed so far, 
and could also contribute significantly to our knowledge on the galactic field 
structure and the galactic CR source distribution.  

\section*{Acknowledgments}
Work supported by CONICET and Fundaci\'on Antorchas, 
Argentina.


\begin{thebibliography}{100}
\bibitem{la83}P.O. Lagage and C.J. Cesarsky, Astron. Astrophys. \textbf{118} (1983) 223;
Astron. Astrophys. \textbf{125} (1983) 249.
\bibitem{bi93}P.L. Biermann, Inv. Rap. High. Papers 45, 
23rd ICRC, Calgary, 1993.
\bibitem{dr94}L.C. Drury et al., Astron. Astrophys. \textbf{287} (1994) 959.
\bibitem{fi86}C.E. Fichtel and J. Linsley, ApJ \textbf{300} (1986) 474.
\bibitem{jo86}J.R. Jokipii and G.E. Morfill, ApJ \textbf{312} (1986) 170.
\bibitem{ko00}K. Kobayakawa, Y.S. Honda, and T. Samura, 
Phys. Rev. D \textbf{66} (2002) 083004.
\bibitem{ka93}S. Karakula and W. Tkaczyk, Astrop. Phys. \textbf{1} (1993) 229.
\bibitem{ca02a}J. Candia, L.N. Epele, and E. Roulet, 
Astrop. Phys. \textbf{17} (2002) 23. 
\bibitem{er97}A.D. Erlykin and A.W. Wolfendale, 
Astrop. Phys. \textbf{7} (1997) 203; Astrop. Phys. \textbf{10} (1999) 69.
\bibitem{sy71}S.I. Syrovatsky, 
Comm. Astrophys. Space Phys. \textbf{3} (1971) 155.
\bibitem{wd84}J. Wdowczyk and A.W. Wolfendale, 
J. Phys. G \textbf{10} (1984) 1453.
\bibitem{pt93}V.S. Ptuskin et al., Astron. Astrophys. {\bf 268} (1993) 726.
\bibitem{ca02b}J. Candia, E. Roulet and L.N. Epele, 
J. High Energy Phys. {\bf 12} (2002) 033.
\bibitem{sw02}S.P. Swordy et al., Astrop. Phys. {\bf 18} (2002) 129. 
\bibitem{sw00}S.P. Swordy and D.B. Kieda, 
Astrop. Phys. \textbf{13} (2000) 137.
\bibitem{ka99}K.-H. Kampert et al., KASCADE Collaboration, OG. 1.2.11, 
26th ICRC, Salt Lake City, 1999; ibidem, astro-ph/0102266.
\bibitem{ca02c}J. Candia, S. Mollerach, and E. Roulet,
J. High Energy Phys. {\bf 12} (2002) 032.
\bibitem{kr65}R.H. Kraichnan, Phys. Fluids {\bf 8} (1965) 1385.
\bibitem{te01}M. Teshima et al., 27th ICRC, Hamburg, 2001.
\bibitem{be90}V.S. Berezinskii et al., {\it Astrophysics of
Cosmic Rays}, Amsterdam: North Holland, 1990.
\bibitem{cl96}R.W. Clay and G.K. Smith,
Pub. Astron. Soc. Aust. {\bf 13} (1996) 121.
\bibitem{ha02}J.L. Han et al., astro-ph/0211197, to appear in {\it Radio Pulsars}, APS Conf. Ser., 
edited by M. Bailes et al. 
\bibitem{wi98}B. Wiebel-Sooth and P.L. Biermann, 
Astron. and Astrophys. \textbf{330} (1998) 389.
\bibitem{ho02}J.R. H\"orandel, Astrop. Phys. \textbf{19} (2003) 193.
\bibitem{na00}M. Nagano and A.A. Watson, Rev. of Mod. Phys. {\bf 72} (2000) 689.
\bibitem{hi84}A.M. Hillas, An. Rev. Astron. Astrophys. {\bf 22} (1984) 425.  
\bibitem{do84}L.I. Dorman, A. Gkhosh and V.S. Ptuskin, 
Pis'ma Astron. Zh. {\bf 10} (1984) 827 [Sov. Astron. Lett. {\bf 10} (1984) 345]. 
\end{thebibliography}
\end{document}